\documentclass[]{article}
\usepackage[a4paper, total={16cm, 24cm}]{geometry}

\usepackage{url} %this package should fix any errors with URLs in refs.
\usepackage{lineno}
\usepackage{soul}

\usepackage{graphicx}
\usepackage{newtxtext}
\usepackage{newtxmath}
\usepackage{natbib}
\usepackage{hyperref}
\usepackage{relsize}
\usepackage{authblk}
\usepackage[symbol]{footmisc}

\DeclareMathAlphabet{\pazocal}{OMS}{zplm}{m}{n}

\renewcommand{\Pr}{{\pazocal{P}{\hskip-0.1mm}r}}
\newcommand{\Pm}{{\pazocal{P}_{\textrm{\scriptsize m}}}}
\newcommand{\Ek}{\pazocal{E}}                         %Ekman number
                       
\newcommand{\Em}{{\pazocal{E}_{\textrm{\scriptsize m}}}}    %Magnetic Ekman number
\newcommand{\Ra}{{\pazocal{R}a}}

\newcommand{\RaR}{{\pazocal{R}}}

\newcommand{\Ratil}{\widetilde{\Ra}}
\newcommand{\drm}{\textrm{d}}

\begin{document}

\title{Scaling of strong-field spherical dynamos}

%%%%%%%%%%%%%%%%%%%%%%%%%%%%%%%%%%%%%%%%%%%%%%%
%
%  AUTHORS AND AFFILIATIONS
%
%%%%%%%%%%%%%%%%%%%%%%%%%%%%%%%%%%%%%%%%%%%%%%%

\author[1]{Robert J. Teed\footnote{Robert.Teed@glasgow.ac.uk}}
\author[2]{Emmanuel Dormy\footnote{Emmanuel.Dormy@ens.fr}}

\affil[1]{{\small School of Mathematics \& Statistics, University of Glasgow, Glasgow G12 9EH, United Kingdom}}
\affil[2]{{\small Département de Mathématiques et Applications, UMR-8553, École Normale Supérieure, CNRS, PSL University, 75005, Paris, France}}
%(repeat as many times as is necessary)

%\correspondingauthor{Robert Teed}{Robert.Teed@glasgow.ac.uk}
%\correspondingauthor{Emmanuel Dormy}{Emmanuel.Dormy@ens.fr}

\maketitle

%
%%%%%%%%%%%%%%%%%%%%%%%%%%%%%%%%%%%%%%%%%%%%%%%

\begin{abstract}
% 150 words (currently 150!)
Numerical experiments of dynamo action designed to understand the generation of 
Earth's
magnetic field produce different regime
branches identified within bifurcation diagrams.
Notable are distinct branches where the resultant
 magnetic field is either weak or strong.
Weak-field solutions are identified by the prominent role of
viscosity (and/or inertia) on the motion, whereas the magnetic field has a leading-order effect on the flow in strong-field solutions.
We demonstrate the persistence of the strong-field branch,
preserving the expected force balance of Earth's core, and provide scaling laws governing its onset as parameters move toward values appropriate for the Geodynamo.
We introduce a new output parameter, based on dynamically important parts of rotational and magnetic forces, that captures expected $O(1)$ values of strong-field solutions throughout input parameter space.
This new measure of the field strength and our bounds on scaling laws
can guide future studies in locating strong-field dynamos in parameter space.

\end{abstract}

%%%%%%%%%%%%%%%%%%%%%%%%%%%%%%%%%%%%%%%%%%%%%%%
%
%  BODY TEXT
%
%%%%%%%%%%%%%%%%%%%%%%%%%%%%%%%%%%%%%%%%%%%%%%%

\section{Introduction}
\label{sec:intro}

The inability to sample conditions within Earth's core directly means numerical simulations can provide an essential
role in progressing our understanding of the Geodynamo mechanism (as well as dynamos of other planets). Varying the key control parameters of the system ($\Ek$, $\Em$, $q$, and $\Ratil$ - each is introduced formally later) maps out part of the solution space. Yet, due to a lack of computational resolution, the regimes identified do not operate under conditions of the core. In particular, viscosity must be set many orders of magnitude too large. Key to making progress is identifying regions of solution space most appropriate to Earth's core at the moderate input parameters available computationally.

Early pioneering work using simulations \citep{christensen1999,Kutzner02} discovered two distinct regimes of Geodynamo action in the accessible parameter space. Dipolar solutions were found at the onset of dynamo action (above the onset of convection), transitioning to a multipolar regime at higher convective driving. Later work \citep{Morin09}
focussed on bifurcations in $\Ratil$-space and showed the existence of subcritical bifurcations and isola. A systematic route to studying subcritical dynamos has recently been introduced by 
\cite{Mannix2022}. More recent work identified different branches of dynamo action solely within the dipolar regime.
It was shown that a bifurcation diagram with two dipolar branches is possible in which the dynamo onsets on a weak-field dipolar branch before transitioning to a strong-field dipolar branch rather than a multipolar branch \citep{dormy2016}. For a range of $\Ratil$, the weak- and strong-field branches are bistable resulting in a double fold and a cusp singularity in the parameter space \citep[see][for a recent review of these ideas]{Dormy2025}. \cite{dormy2018} then considered lower $\Ek$; however, with no tracking of bifurcations, the reported strong-field dynamos were based on unconfirmed scaling laws, not on observation of separate branches and bistability. 
\cite{menu2020} further showed that the strong-field solution exists at a level of forcing for which the weak-field destabilises to a multipolar state. Recently, \cite{Guseva2025}
were able to provide a mathematical framework for studying the destabilisation of the weak-field solution to the strong-field regime.

The importance of the strong-field branch lies in its solutions providing an anticipated balance of forces for Earth's core, in contrast to the other branches discussed. MAC-balance, between magnetic (i.e.~Lorentz), Archimedean (i.e. buoyancy), and Coriolis forces, is expected to prevail in Earth's core \citep[see also discussions in][]{soderlund2012,soderlund2015,buffett2000,mason2022,horn2022,horn2025}. 
The strong-field regime is thus a part of solution space requiring further investigation to check its existence and scaling as parameters are gradually moved towards those of Earth's core.

In this work we demonstrate the persistence of the strong-field branch as $\Ek$ and $\Em$ are lowered toward values appropriate to the Geodynamo. We investigate scaling laws, as a function of $\Ek$, for the critical value of $\Em$ below which distinct
weak- and strong-field branches can be found. Furthermore, we introduce an alternative definition of the Elsasser number based on the ratio of \emph{curls} of the Lorentz and Coriolis forces. In contrast to previous definitions, this new output parameter is able to capture the expected $O(1)$ values for strong-field solutions and maintain this across a wide range of values of $\Ek$.

%%%%%%%%%%%%%%%%%%%%%

\section{Methods}

\subsection{Governing equations}

The system under consideration is a spherical shell (bounded between $r=r_\text{i}$ and $r=r_\text{o}$ with aspect ratio, $\chi=r_\textrm{i}/r_\textrm{o}$) filled with a conducting Boussinesq fluid and rotating with rate $\Omega$ about the vertical. The variables are nondimensionalised using lengthscale, $d=r_\textrm{o}-r_\textrm{i}$, timescale, $d^2/\eta$, temperature scale, $\Delta T$, and magnetic scale $\sqrt{\mu_0\rho\eta\Omega}$. The evolution of the velocity of the fluid, $\mathbf{u}$, the temperature $T$, and the magnetic induction, $\mathbf{B}$, are governed by the following set of non-dimensionalised equations
\begin{subequations}
\label{eq:gov}
\begin{gather}
 \Em\left(\frac{\partial\mathbf{u}}{\partial t} + \mathbf{u}\cdot\nabla\mathbf{u}\right) = -\nabla p - 2\, \mathbf{\hat{z}}\times\mathbf{u} + (\nabla\times\mathbf{B})\times\mathbf{B} + \widetilde{\Ra}\, T\, \mathbf{r} + \Ek\, \nabla^2\mathbf{u}\,, \label{eq:NSeq} \tag{\theequation $a$}
 \\
 \frac{\partial T}{\partial t} + \mathbf{u}\cdot\nabla T = q\, \nabla^2T\,, \qquad \qquad
 \frac{\partial\mathbf{B}}{\partial t} - \nabla\times(\mathbf{u}\times\mathbf{B}) = \nabla^2\mathbf{B}\,, \label{eq:indeq}\tag{\theequation $b,c$} \\
 \nabla\cdot\mathbf{u} = 0\,, \qquad
 \nabla\cdot\mathbf{B} = 0\,,\tag{\theequation $d,e$}
\end{gather}
\end{subequations}
where $p$ is the pressure and $\mathbf{r}=(1-\chi)\, r \,\mathbf{e}_r\, .$
The nondimensional parameters in equations (\ref{eq:gov})
are the Ekman number, $\Ek$, magnetic Ekman number, $\Em$, modified Rayleigh number, $\Ratil$, and Roberts number, $q$,  defined as
\begin{subequations}
\begin{equation}
 \Ek=\frac{\nu}{\Omega d^2}\, ,
 \qquad \Em=\frac{\eta}{\Omega d^2}\, ,
 \qquad \widetilde{\Ra}=\frac{\alpha g_\textrm{o}\Delta T d}{\Omega\eta}\, ,  
 \qquad q=\frac{\kappa}{\eta}\, .
 \tag{\theequation $a$--$d$}
\end{equation}
\end{subequations}
The Rayleigh number used in this work is related to the `rotational' Rayleigh number, $\Ra=\alpha g_\textrm{o}\Delta T d/\Omega\kappa$, and the classical (non-rotational) Rayleigh number, $\RaR=\alpha g_\textrm{o}\Delta T d^3/\nu\kappa$, through $\Ratil=q\Ra = q\Ek \RaR$. For convenience, the Rayleigh number is often measured by its supercriticality, $\widetilde{\Ra}'=\Ratil/q\Ra_c=\Ra/\Ra_c$, where $\Ra_c$ is the value of $\Ra$ for the onset of non-magnetic convection (for a given $\Ek$).

We focus on a set-up broadly appropriate for the Geodynamo (although the results that follow also have relevance to other natural dynamos).
We therefore set $\chi=0.35$ and choose rigid, electrically-insulating, isothermal boundary conditions with differential heating. 
The governing equations, (\ref{eq:gov}$a$--$e$), (subject to our boundary conditions) are solved using the Leeds spherical dynamo code \citep{willis2007}.

\subsection{Elsasser number and Curlsasser number}

The Elsasser number, $\Lambda$, is a measure of the ratio of the Lorentz force to the Coriolis force. It is often introduced to determine whether a particular dynamo solution has a `strong' magnetic field with $\Lambda\sim \textit{O}(1)$, suggestive of a dominant balance between Lorentz and Coriolis forces as part of a wider MAC-balance.

In its broadest definition the Elsasser number can be written as 
\begin{equation}
\Lambda = \frac{\{\mathbf{F}_L\}}{\{\mathbf{F}_C\}} = \frac{\{(\mu\rho)^{-1}(\nabla\times\mathbf{B})\times\mathbf{B}\}}{\{2\Omega\hat{\mathbf{z}}\times\mathbf{u}\}} = \frac{B^2}{2\mu\rho\Omega U\ell_B},
\label{eq:Elsasserprimitive}
\end{equation}
where $\mathbf{F}_L$ and $\mathbf{F}_C$ are the Lorentz and Coriolis forces, respectively, and $\{\cdot\}$ is an `order-of-magnitude' operator indicating some process of approximating typical values and averaging. In (\ref{eq:Elsasserprimitive}), $U$ and $B$ are typical (e.g. rms) values of the velocity and magnetic field, respectively and $\ell_B$ is a lengthscale appropriate for the magnetic field.

One must then make choices for measures of $U$, $B$, and $\ell_B$, leading to two common versions of the Elsasser number:
\begin{align}
\Lambda &= \frac{B^2}{2\mu\rho\Omega\eta} &&\text{ for } U\sim\frac{\eta}{d}, \text{ with } \ell_B\sim d, \label{eq:Elsasserclassical} \\
\Lambda'&= \frac{B^2d}{2\mu\rho\Omega\eta Rm \ell_B} = \Lambda\frac{d}{Rm\ell_B} &&\text{ for } U\sim\frac{Rm}{d/\eta}. \label{eq:Elsassermodified}
\end{align}
Here, $Rm$ %=Ud/\eta$
is the magnetic Reynolds number based on the rms value of the velocity
and\\
$\ell_B = \int_V\mathbf{B}^2 \drm V / \int_V (\nabla\times\mathbf{B})^2 \drm V $
is a measure of the magnetic dissipation lengthscale.
The classical definition of the Elsasser number is given by (\ref{eq:Elsasserclassical}) whereas $\Lambda'$ is a `modified' \citep{dormy2016} or `dynamic' \citep{soderlund2012} Elsasser number. 

The leading order balance of forces is however partly disguised by the large pressure gradient force. Recent work \citep{teed2023} has shown that the leading order balance of forces can be readily recovered by filtering out dynamically unimportant gradient parts of all forces, forming solenoidal forces.
If the purpose of the Elsasser number is to indicate balance between Lorentz and Coriolis forces, another logical option for quantifying the influence of the Lorentz force is to take the ratio of the \emph{curls} of the original forces. We therefore introduce {a new, previously unused measure,} the `Curlsasser number'\footnote{The symbol $\sigma$ chosen here for its resemblance to a curling stone!},
given by
\begin{equation}
\sigma = \frac{\{\mathbf{C}_L\}}{\{\mathbf{C}_C\}} = \frac{\{(\mu\rho)^{-1}\nabla\times((\nabla\times\mathbf{B})\times\mathbf{B})\}}{\{2\Omega\nabla\times(\hat{\mathbf{z}}\times\mathbf{u})\}} = \frac{B^2\ell_z}{2\mu\rho\Omega U\ell_B\ell_C},
\label{eq:Curlsasser}
\end{equation}
where $\ell_z$ is the vertical lengthscale of the flow and $\ell_C$ is some correlation lengthscale coupling the magnetic field and its curl.
In practice the quantity on the RHS of (\ref{eq:Curlsasser}) will be difficult to calculate because of the uncertainty in measuring $\ell_C$, although, in a similar vein to $\ell_B$, one could approximate
$\ell_z = \int_V(\mathbf{\hat{z}\times\mathbf{u}})^2 \drm V / \int_V (\nabla\times(\hat{\mathbf{z}}\times\mathbf{u}))^2 \drm V$.

The uncertainties in the definitions of the Elsasser numbers given by (\ref{eq:Elsasserprimitive}) and (\ref{eq:Curlsasser}) can be avoided by calculating the relevant forces and/or curls of forces directly. In this work we therefore present $\Lambda'_\textrm{F}$ based on the ratio of the directly calculated forces and $\sigma$ based on the ratio of the directly calculated curls of forces; i.e.~we set:
\begin{subequations}
\label{eq:newelsasser}
\begin{equation}
\Lambda'_\textrm{F}=\sqrt{\frac{\int_V((\nabla\times\mathbf{B})\times\mathbf{B})^2\drm V}{\int_V(2\hat{\mathbf{z}}\times\mathbf{u})^2\drm V}},
\qquad
\sigma=\sqrt{\frac{\int_V(\nabla\times((\nabla\times\mathbf{B})\times\mathbf{B}))^2\drm V}
{\int_V(2\nabla\times(\hat{\mathbf{z}}\times\mathbf{u}))^2\drm V}}.\tag{\theequation $a,b$}
\end{equation}
\end{subequations}
The measures given by (\ref{eq:newelsasser}) provide a direct estimate of the relevant force ratio unlike those given by (\ref{eq:Elsasserclassical}-\ref{eq:Elsassermodified}), which rely on  choices being made for the relevant lengthscales and flow/field magnitudes.
Both $\Lambda'_\text{F}$ and $\sigma$ are derived from measurable quantities and can be readily calculated from simulation data at no extra computational cost compared to $\Lambda$ and $\Lambda'$. 
Indeed, \cite{soderlund2015} calculated the ratio of the forces (i.e.~$\Lambda'_\text{F}$) and compared various other definitions of the Elsasser number against this.
While our primary focus is on the new measure, $\sigma$, we also present $\Lambda'$ and $\Lambda'_\text{F}$ for comparison with previous studies.

%%%%%%%%%%%%%%%%%%%%%%%%%
\section{Strong-field dynamos at low $\Ek$}

We have performed a large simulation suite, involving significant computational resources, to focus on tracking bifurcations and weak- and strong-field branches in $\Ratil$-space as $\Ek$ and $\Em$ vary. The value of $q$, whilst allowed to vary, does so in a particular manner: it is set to  $q= \Ek/\Em$ in all simulations.
Previous work in the community has tended to use $\Pr=\nu/\kappa=\Ek/(q\Em)$ and $\Pm=\nu/\eta=\Ek/\Em$ as input parameters (rather than $q$ and $\Em$); in terms of these controlling numbers 
our simulation suite can be equivalently thought of as having set $\Pr=1$ (and thus $\Pm=q$) in all simulations. We track bifurcations as a function of supercriticality, $\Ratil'$; the values of $\Ra_c$ used in our study are listed in Table \ref{tab:rac}.

\begin{table}
  \begin{center}
\def~{\hphantom{0}}
  \begin{tabular}{c|c|cc|ccc}
      $\Ek$  & $\Ra_c$  & $\Em_c$ & $\Pm_c$ & $\Em_c^\textrm{S}$ & $\Pm_c^\textrm{S}$  \\[3pt]
     $3\times10^{-4}$ &  60.8 & $2\times10^{-4}$ & $1.5$ ($\sim1.5$) & $2.1\times10^{-5}$ & 14 \\
       $10^{-4}$ &  69.7 & $2\times10^{-4}$ & $0.5$ ($\sim0.45$) & $2\times10^{-5}$ & 5 \\
       $3\times 10^{-5}$ & 84.9 & $1.2\times10^{-4}$ & $0.25$ ($\sim0.18$) & $6\times10^{-6}$ & 5 \\
       $10^{-5}$ &  106 & $5\times10^{-5}$ &  $0.2$ ($\sim0.09$) & $5\times10^{-6}$ & 2 \\
       $10^{-6}$ &  179 & $3.3\times10^{-5}$ & $0.03$ ($\sim0.045$) & $10^{-6}$ & 1 \\
  \end{tabular}
  \end{center}
  \caption{Critical values of key parameters for values of the Ekman number considered in this study (all for $\Pr=1$). The (rotational) critical Rayleigh number, $\Ra_c$, is for the onset of (non-magnetic) convection.
  No superscript indicates (approximate) maximal $\Em$ values (minimal $\Pm$) required for weak-field dipolar dynamo action.
The superscript `S' indicates equivalent quantities for the strong-field dipolar branch.
}
\label{tab:rac}
\end{table}

\subsection {Bifurcation diagrams using the modified Elsasser number}

\begin{figure}[hb]
\centerline{
(a)\hskip 0mm
{\includegraphics[width=0.32\linewidth]{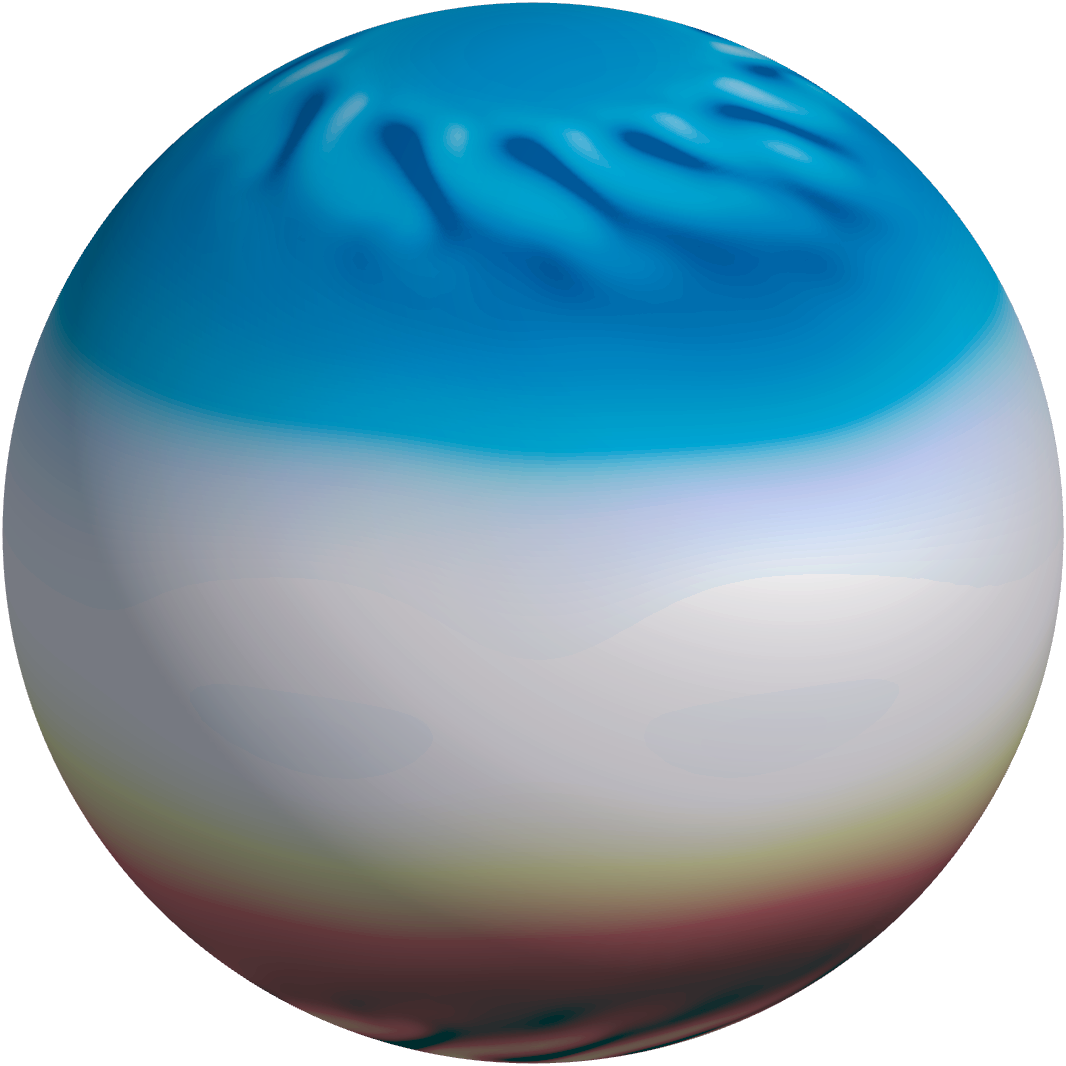}}
\hspace{5mm}
(b)\hskip-0mm
{\includegraphics[width=0.32\linewidth]{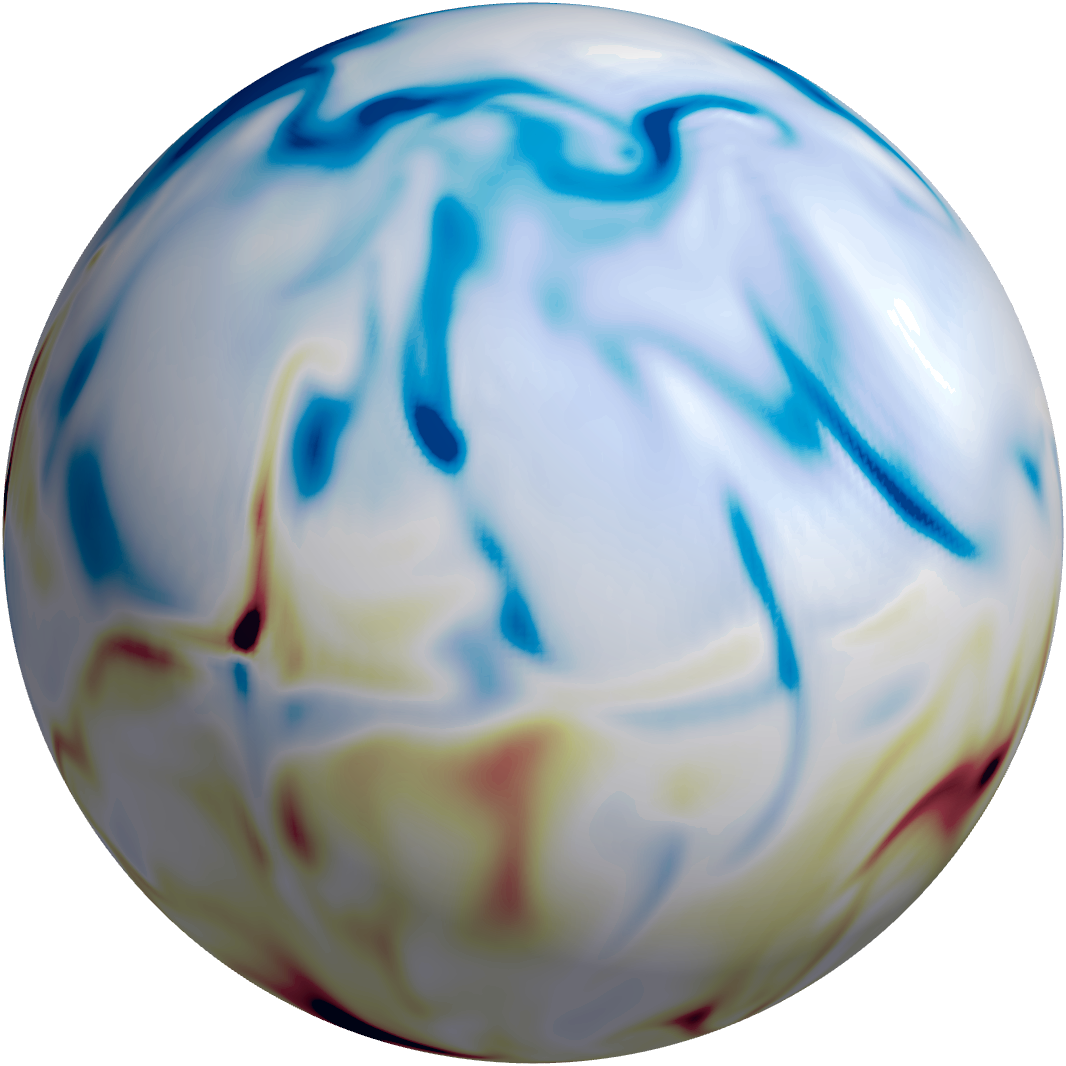}}}
\centerline{(c)\hskip 0mm
{\includegraphics[width=0.32\linewidth]{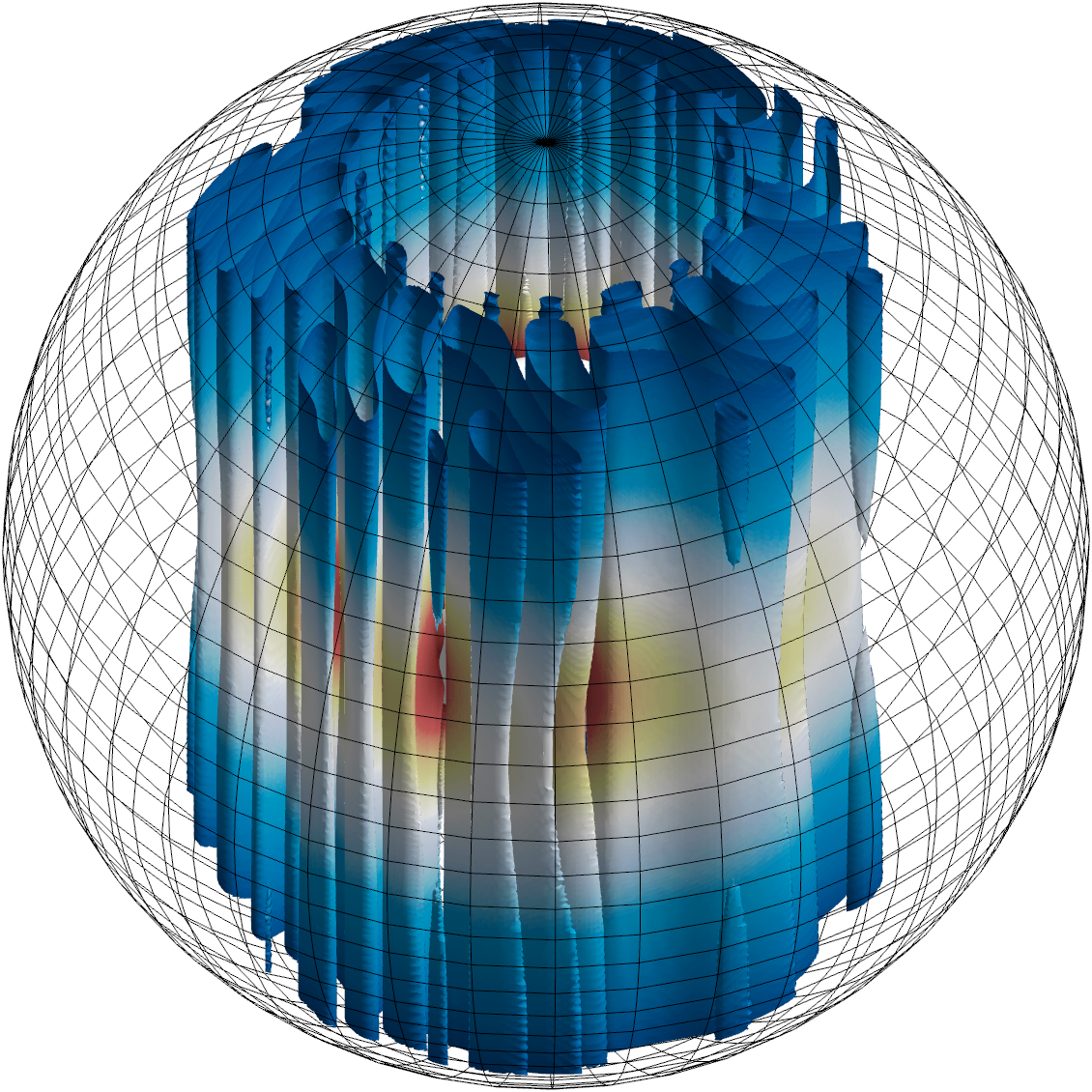}}
\hspace{5mm}
(d)\hskip -0mm{\includegraphics[width=0.32\linewidth]{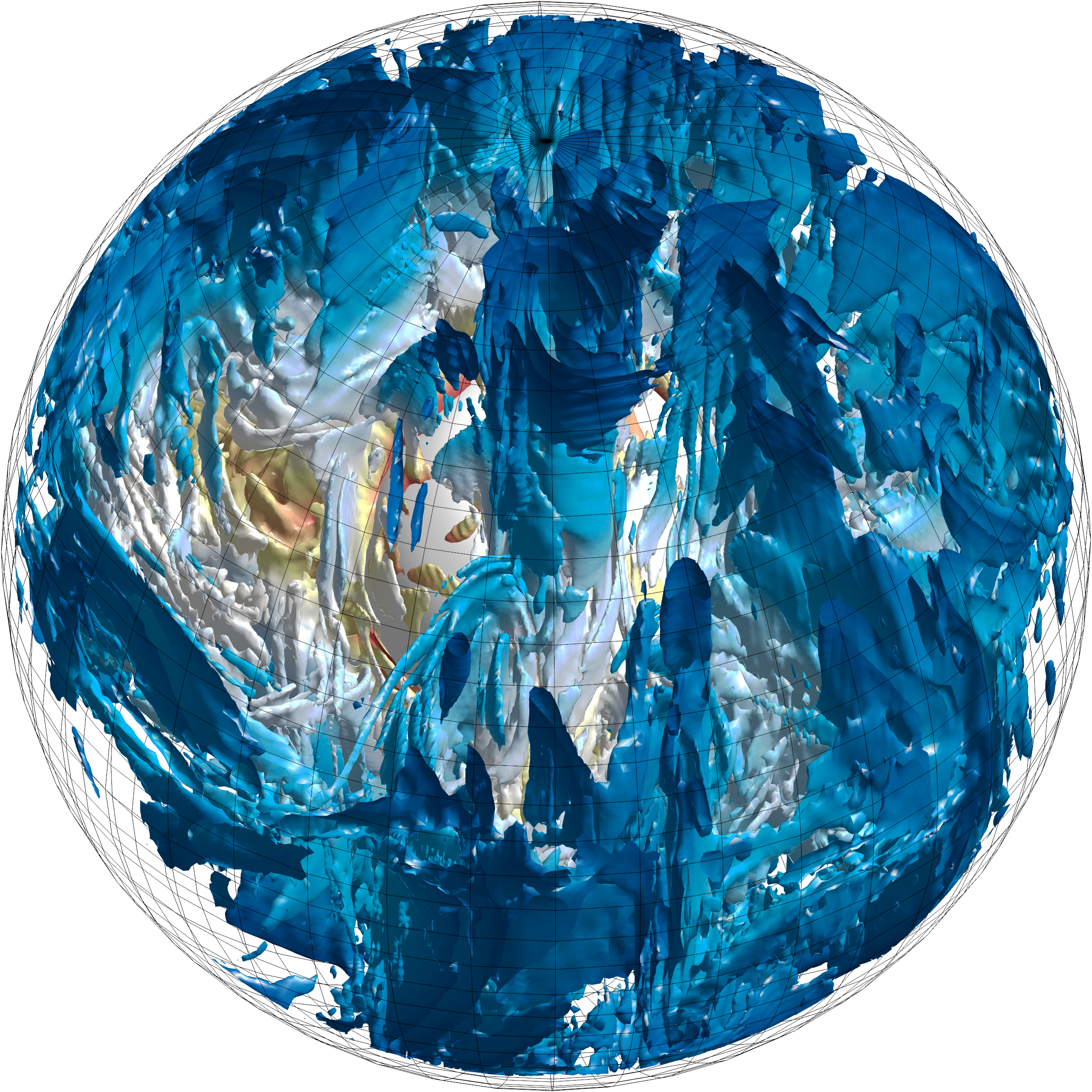}}
}

\caption{Snapshots of the radial magnetic field, $B_r$, on a spherical surface at the outer boundary for: (a) a weak-field solution ($\Ratil'=3$, $\Ek=10^{-5}$, $\Em=2\times10^{-6}$);
(b) a strong-field solution ($\Ratil'=5$, $\Ek=10^{-5}$, $\Em=2\times10^{-6}$).
(c-d) As (a-b) but displaying isocontours of the vertical vorticity, $\omega_z$, coloured with the temperature field, $T$. 
}
\label{fig:3DWFSF}
\end{figure}

\begin{figure}[ht]
\centerline{
(a)\hskip -3mm
{\includegraphics[page=1,width=0.47\linewidth]{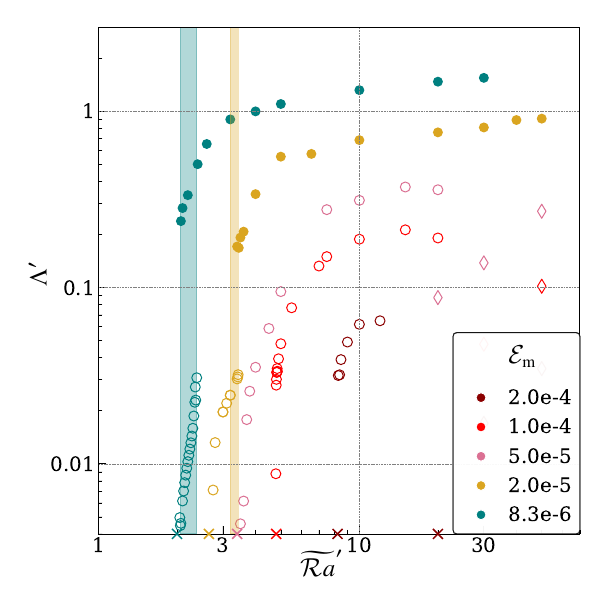}
}
\hspace{1mm}
(b)\hskip -3mm
{
\includegraphics[page=2,width=0.47\linewidth]{figures/RatpElp_plots.pdf}
}
\hspace{1mm}
}
\centerline{
(c)\hskip -3mm
{
\includegraphics[page=3,width=0.47\linewidth]{figures/RatpElp_plots.pdf}
}
\hspace{1mm}
(d)\hskip -3mm
{
\includegraphics[page=4,width=0.47\linewidth]{figures/RatpElp_plots.pdf}
}
}
\vspace{-2mm}
\caption{Modified Elsasser number, $\Lambda'$, as a function of supercriticality, $\Ratil'$, for (a) $\Ek=10^{-4}$, (b) $\Ek=3\times10^{-5}$, (c) $\Ek=10^{-5}$, and (d) $\Ek=10^{-6}$. Each point represents a single simulation where data has been volume averaged and time-averaged.
Filled/unfilled circles respectively indicate strong-/weak-field dipolar; diamonds for fluctuating multipolar; cross symbols for failed dynamos (with data points moved to the lower horizontal axis).
Shaded regions 
indicate observed
bistability between the weak- and strong-field branches.
}
\label{fig:Elsasserbifurcation}
\end{figure}

Bifurcations in $\Ratil'$-$\Lambda'$-space are shown in Fig.~\ref{fig:Elsasserbifurcation}, where each panel displays averaged simulation data at a different value of $\Ek$.
In each plot we find weak-field solutions (with small $\Lambda'$) above some critical value of $\Ratil'$ below which dynamo action is lost.
However, such solutions only exist if $\Em<\Em_c$ ($\Em_c$ being the onset value for dynamo action; see Table \ref{tab:rac}, where values in parentheses are estimated from Fig.~1 of \cite{christensen2006}). If $\Em$ is small enough the weak-field solution becomes unstable (as $\Ratil$ is increased) leading to a transition to a separate strong-field branch characterised by large $\Lambda'$, with bistability between the branches.
It is therefore immediately apparent from the plots of Fig.~\ref{fig:Elsasserbifurcation} that the general picture identified by \cite{dormy2016}
remains valid as the Ekman number is reduced. That is, for a given $\Ek$, separate weak- and strong-field branches can only be found for small enough $\Em$.
On both the weak- and strong-field branch the surface magnetic field is dominantly dipolar
(see Fig.~\ref{fig:3DWFSF}(a-b)) but the flow differs drastically on the two branches (see Fig.~\ref{fig:3DWFSF}(c-d)).

For each $\Ek$, it is possible to approximate a critical value of the magnetic Ekman number, $\Em_c^\text{S}$,
below which the strong-field branch exists (or, equivalently, for $\Pm>\Pm_c^\text{S}$). For the purposes of our numerical work we define $\Em_c^\text{S}$ to be the largest value of $\Em$ for which bistability is found (see Table \ref{tab:rac}).
Tracking of branches at further values of $\Em$ would be required to improve the accuracy of these critical values.

For a given $\Ek$, both the separation between the branches in $\Lambda'$-space and the range of $\Ratil'$ for which bistability is observed reduces as $\Em$ increases. This continues until the branches merge and the double turning point unwinds at $\Em\sim\Em_c^\text{S}$, marking the point in $\Em$-space above which the strong-field branch no longer exists. These properties are supportive of the existence of a cusp singularity \citep{dormy2016}. Separately, bistability between the branches also moves to larger $\Ratil$ as $\Em$ increases reflecting the increased forcing required to drive dynamos at higher magnetic diffusivity.

For the values of $\Ek$ tested, the strong-field branch was not observed
for $\Ek<\Em$ (i.e.~for $\Pm<1$) although it does exist for $\Ek=\Em=10^{-6}$ (i.e.~$\Pm=1$). Hence below this value of $\Ek$ the strong-field branch will presumably exist for $\Pm<1$, allowing for its preservation as both $\Em$ and $\Pm$ are reduced to small values appropriate for Earth's core. This further supports the idea of a distinguished limit (see section \ref{sec:dishlim}) proposing $\Em$ (or $\Pm$) should scale with $\Ek$ to maintain strong-field solutions.

An unexpected feature of the results presented in Fig.~\ref{fig:Elsasserbifurcation} is the reduction in $\Lambda'$ as $\Ek$ is lowered. Its value both at the lower end of the strong-field branch and for strong-field solutions more generally is found to drop below an \textit{O}(1) value (particularly so at $\Ek=10^{-6}$). This unanticipated attribute prompts a reassessment of the definitions used for the Elsasser number and introduction of the Curlsasser number.

\subsection {Bifurcation diagrams using the Curlsasser number}

\begin{figure}
\centerline{
(a)\hskip -2mm
{\includegraphics[page=1,width=0.47\linewidth]{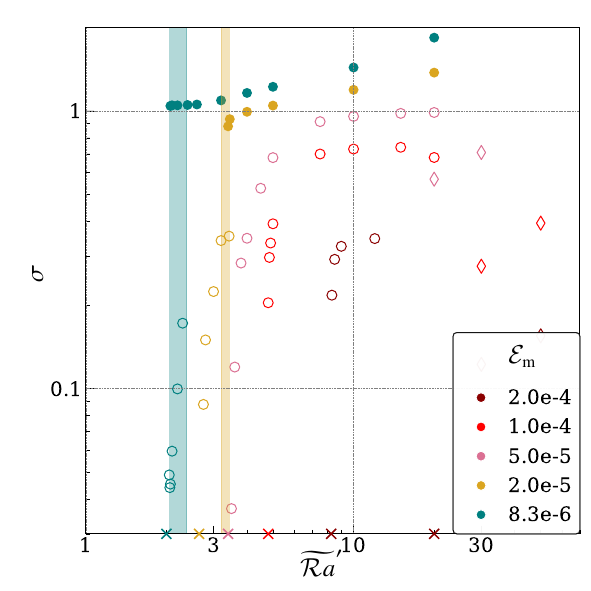}}
\hspace{1mm}
(b)\hskip -2mm
{\includegraphics[page=2,width=0.47\linewidth]{figures/RatpElc_plots.pdf}}
\hspace{1mm}
}
\centerline{
(c)\hskip -2mm
{\includegraphics[page=3,width=0.47\linewidth]{figures/RatpElc_plots.pdf}}
\hspace{1mm}
(d)\hskip -2mm
{\includegraphics[page=4,width=0.47\linewidth]{figures/RatpElc_plots.pdf}}
}
\vspace{-2mm}
\caption{Curlsasser number, $\sigma$, as a function of supercriticality, $\widetilde{\Ra}'$, for (a) $\Ek=10^{-4}$, (b) $\Ek=3\times10^{-5}$, (c) $\Ek=10^{-5}$, and (d) $\Ek=10^{-6}$. 
See Fig.~\ref{fig:Elsasserbifurcation} for explanations of symbols and shaded regions.
}
\label{fig:Curlsasserbifurcation}
\end{figure}

Bifurcations in $\Ratil'$-$\sigma$-space are shown in Fig.~\ref{fig:Curlsasserbifurcation}.
A key difference is found in the magnitude of $\sigma$ when compared with the values of $\Lambda'$ from Fig.~\ref{fig:Elsasserbifurcation}. strong-field solutions cluster around $\sigma=1$. Crucially this remains the case as $\Ek$ or $\Em$ are lowered, in contrast to the behaviour of $\Lambda'$. This suggests that $\sigma$ is a more robust measure of the ratio of the Lorentz to Coriolis effects that remains $O(1)$ as input parameters are moved to more realistic values.

Naturally, the (non-)existence of separate branches and positions of bistability in $\Ratil'$-space are unchanged from Fig.~\ref{fig:Elsasserbifurcation}. However, separation between branches in $\sigma$-space differs from the equivalent in $\Lambda'$-space.

\subsection {Strong-field master curve}

\begin{figure}[h]
\centerline{
(a)\hskip -2mm
{\includegraphics[page=2,width=0.45\linewidth]{figures/RatEl_plots.pdf}}
\hspace{4mm}
(b)\hskip -2mm
{\includegraphics[page=3,width=0.45\linewidth]{figures/RatEl_plots.pdf}}
}
\centerline{
(c)\hskip -2mm
{\includegraphics[page=4,width=0.45\linewidth]{figures/RatEl_plots.pdf}}
\hspace{4mm}
(d)\hskip -2mm
{\includegraphics[width=0.45\linewidth]{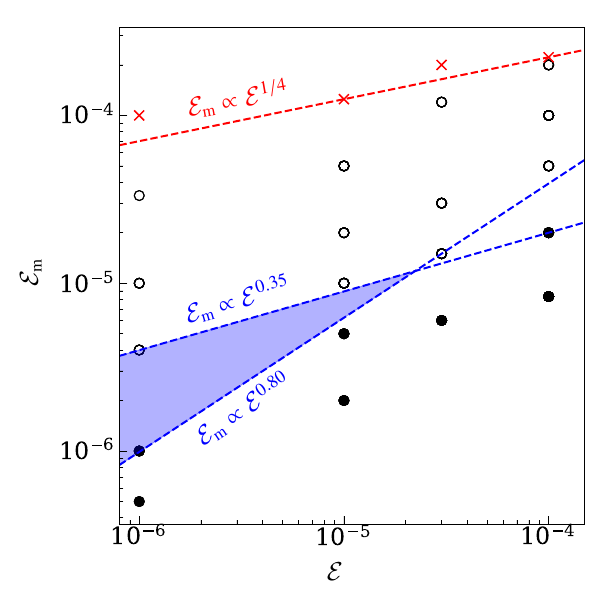}}
}
\vspace{-2mm}
\caption{Dependence on the Rayleigh number, $\Ratil$, of (a) $\Lambda'$, the modified Elsasser number (b) $\Lambda'_\text{F}$, the Elsasser number (directly from forces), and (c) $\sigma$, the Curlsasser number (directly from solenoidal forces). 
(d) Location of strong-field solutions in Ekman number space.
Blue lines show extrema for possible scalings of $\Em$ with $\Ek$ for preservation of strong-field solutions; the red line is the proposed scaling for the onset of the weak-field branch \citep{christensen2006}.
See Fig.~\ref{fig:Elsasserbifurcation} for explanations of symbols.
}
\label{fig:ratilde}
\end{figure}

Data points for all values of Ekman number are presented in Figs.~\ref{fig:ratilde}(a-c), as function of $\Ratil$. Solutions on the strong-field branch are shown as filled symbols whereas all other regimes are displayed with unfilled symbols. {Fig.~\ref{fig:ratilde}(a) is an updated version of Fig.~9(b)} from \cite{dormy2018} now incorporating several values of the Ekman number, across several orders of magnitude. In contrast with that work, a curve of strong-field solutions with $\Lambda'\sim O(1)$ is no longer observed. As discussed earlier, this results from a reduction in $\Lambda'$-space of the location of the strong-field branch as $\Ek$ is reduced.   
Conversely, the Curlsasser number shown in Fig.~\ref{fig:ratilde}(c) captures the strong-field solutions on a striking master curve with $\sigma\sim1$.
The plots of Figs.~\ref{fig:ratilde}(a-c) incorporate simulations across a range of all input parameters, further
highlighting the
robustness of $\sigma$ in identifying strong-field solutions.

Some solutions are `strong-ish' in the sense that they have $\Lambda'$ and/or $\sigma$ close to 1 but are not found above and beyond a turning point in the bifurcation diagrams of Figs.~\ref{fig:Elsasserbifurcation}-\ref{fig:Curlsasserbifurcation}. Such solutions are typically very close to the cusp singularity in parameter space, for which the bistability is replaced by a smooth transition \citep[see discussion in][]{Dormy2025}.
For these dynamos, {the Lorentz force does not enter the primary balance and} inertia and/or viscous effects remain important in the solution (see section \ref{sec:forces}). {Such solutions also destabilise to a multipolar regime at larger convective driving, unlike the genuine strong-branch solutions \citep{menu2020}}.

\subsection{Force balance}
\label{sec:forces}

\begin{figure}
\centerline{
(a)\hskip -5mm
{\includegraphics[width=0.32\linewidth]{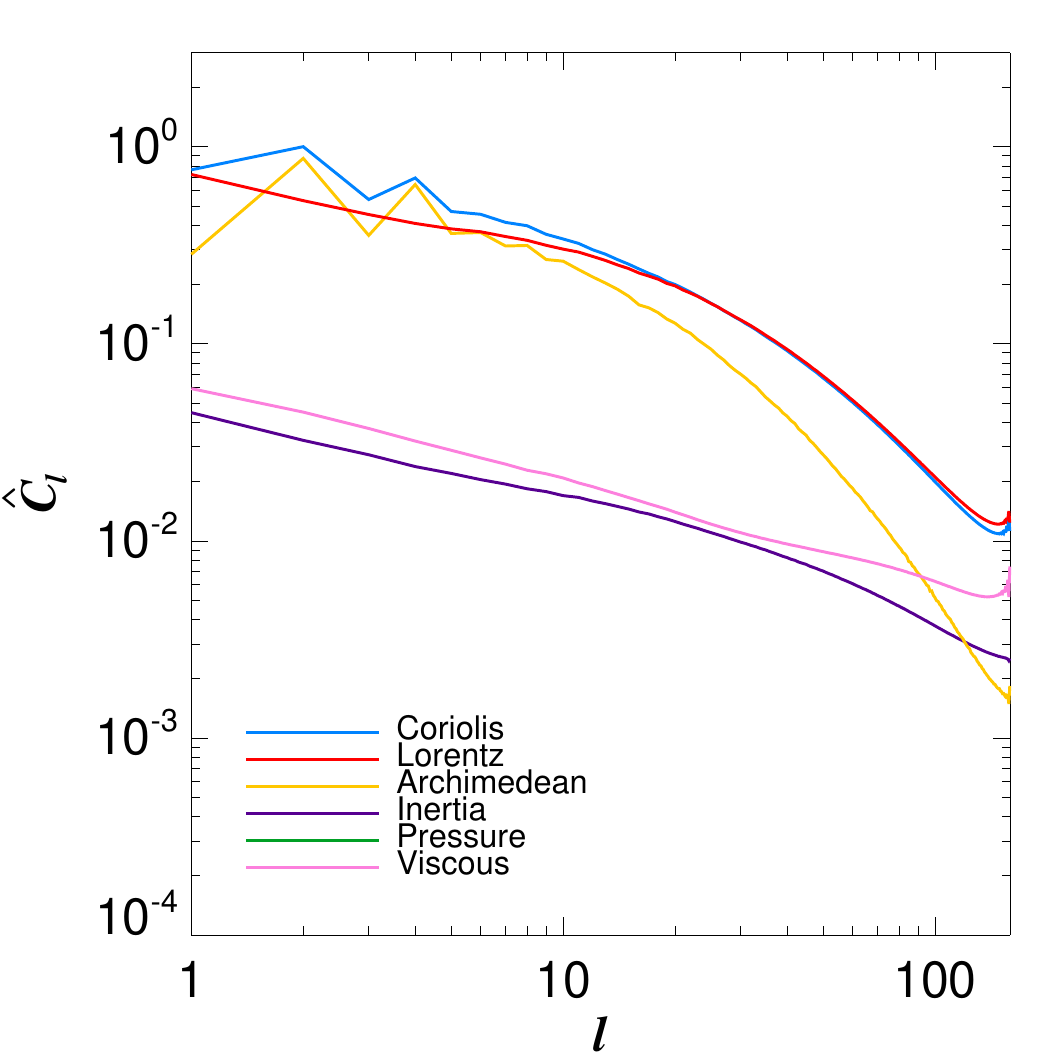}}
(b)\hskip -5mm
{\includegraphics[width=0.32\linewidth]{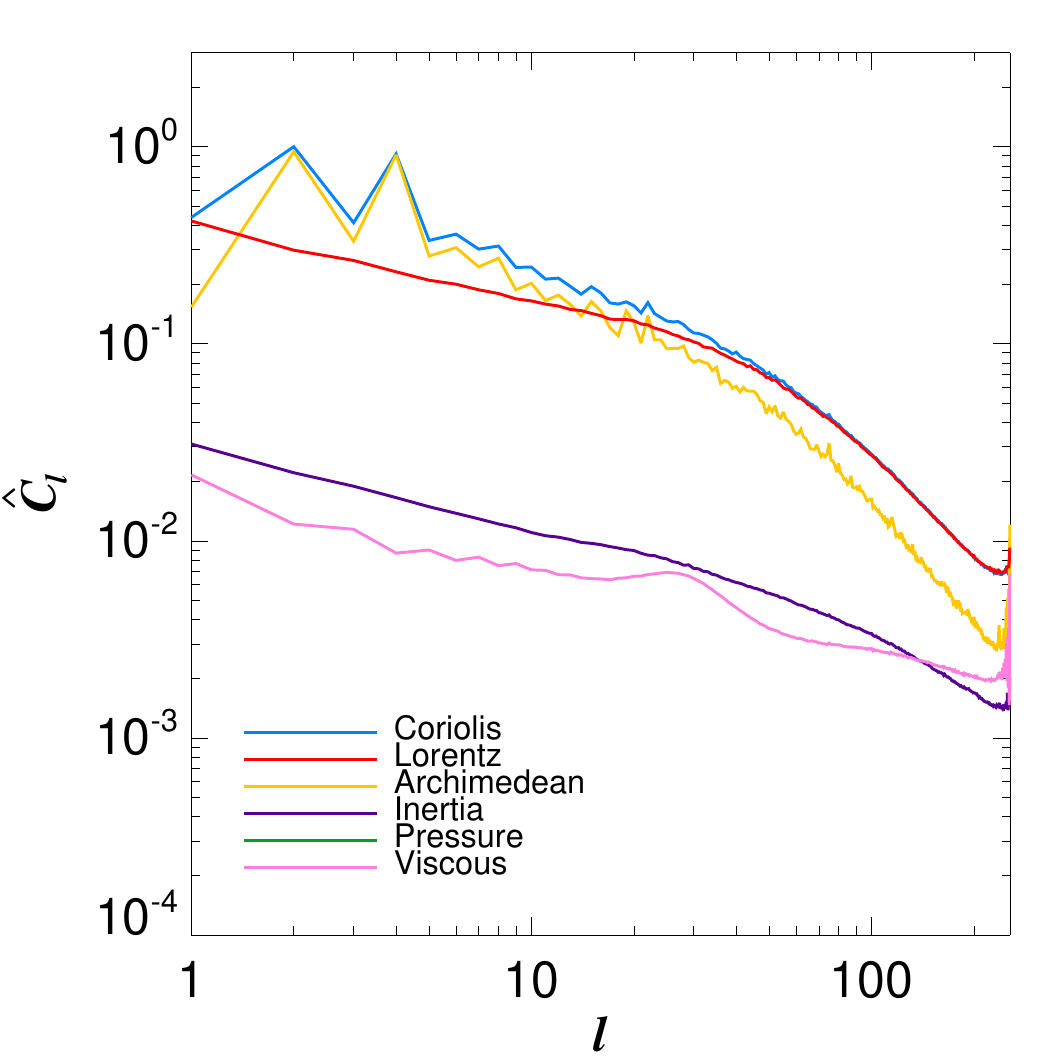}}
(c)\hskip -5mm
{\includegraphics[width=0.32\linewidth]{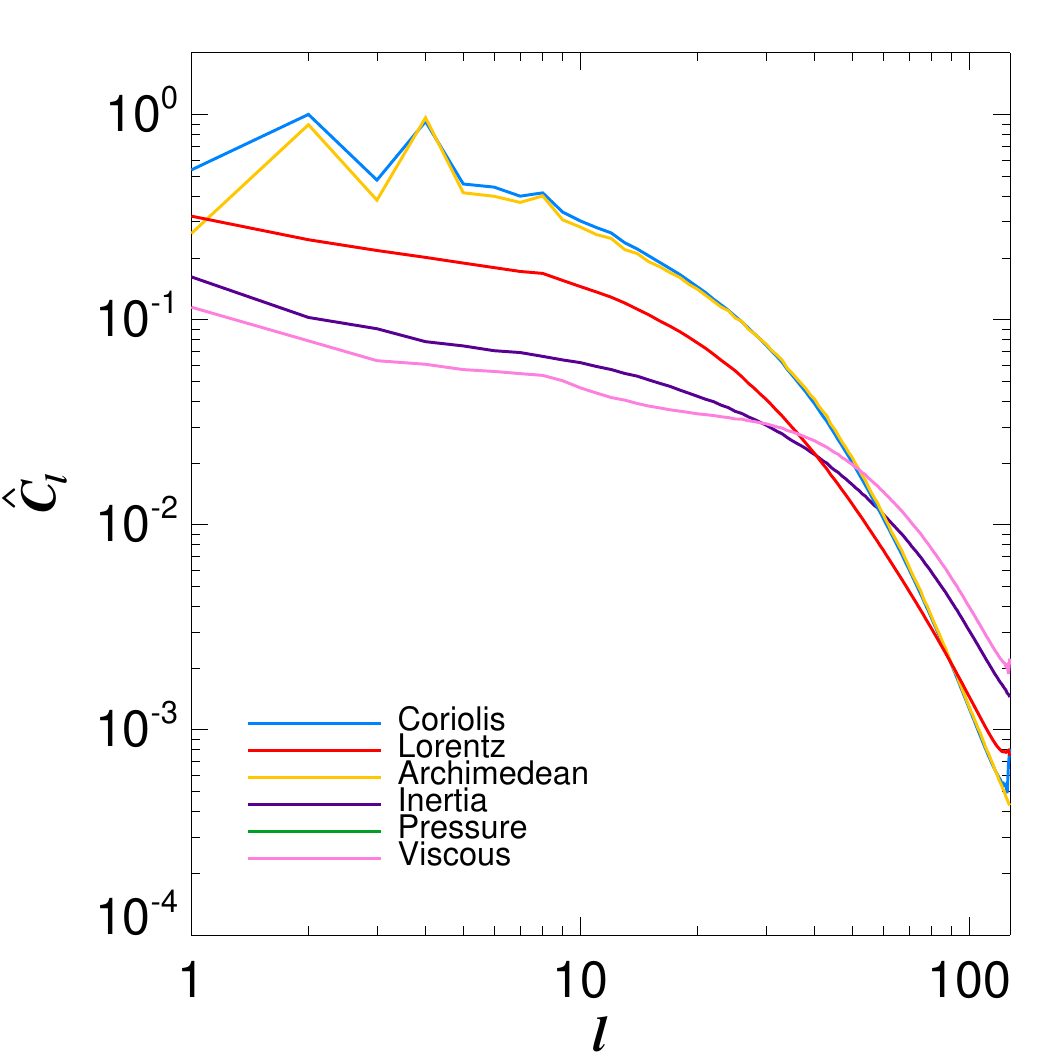}}
\hspace{1mm}
}
\vspace{-2mm}
\caption{Solenoidal forces, $\hat{C}_l$, as a function of lengthscale, $l$ for simulations with: (a) $\Ek=10^{-5}$, $\Em=5\times10^{-6}$, and $\Ratil'=10$ (strong-field); (b) $\Ek=10^{-6}$, $\Em=10^{-6}$, and $\Ratil'=10$ (strong-field); (c) $\Ek=10^{-4}$, $\Em=10^{-4}$, and $\Ratil'=10$ (`strong-ish' field).
}
\label{fig:forces}
\end{figure}

Solutions on the strong-field branch have previously been shown to be in MAC-balance or MC-balance across a range of lengthscales \citep{teed2023} and broadly by position in the domain \citep{dormy2016}. Fig.~\ref{fig:forces} shows how MAC-balance (at the largest lengthscales) and MC-balance (at intermediate to small lengthscales) is preserved on the strong-field branch as $\Ek$ is reduced. Figs.~\ref{fig:forces}(a)-(b) are for simulations chosen at $\Ek=10^{-5}$ and $\Ek=10^{-6}$ to demonstrate the balance at the lowest Ekman numbers available. Similar results are found at larger $\Ek$; indeed, an equivalent plot for $\Ek=10^{-4}$ can be found in Fig.~3(e) of \cite{teed2023}. Conversely, solutions that are `strong-ish' (Fig.~\ref{fig:forces}(c)) have a diminished role for the Lorentz force, no fully formed MAC-balance at large scales, and have a greatly increased role for inertial and/or viscous forces. 

%%%%%%%%%%%%%%%%%%%%%%%%%
\section {Scaling of the strong-field branch with $\Ek$ for a distinguished limit}
\label{sec:dishlim}

The data presented in our study allows for an estimation of a relationship between $\Ek$ and $\Em$ for the preservation of strong-field solutions with MAC-balance in the limit of small $\Ek$. Following \cite{dormy2016}, we relate the Ekman numbers through a single small parameter, $\varepsilon$, by setting $\Ek=\Ek_0\varepsilon$ and $\Em=\Em_0\varepsilon^\alpha$ for some constants $\Ek_0$, $\Em_0$, and $\alpha$, to be determined. 
The values of $\Ek_0$ and $\Em_0$ are selected to set the desired regime in parameter space and $\alpha$ sets the scaling relationship to preserve the desired behaviour as $\varepsilon\to0$. Alternatively, one can pose the same scaling using $\Pm$, in which case $\Pm=\Ek/\Em=\Pm_0\varepsilon^{(1-\alpha)}$ where $\Pm_0=\Ek_0/\Em_0$.

Assuming that our values of $\Ek$ and $\Em$ are approaching some asymptotic behaviour, the values of $\Em_c^\text{S}$ (Tab.~\ref{tab:rac}) offer a range of admissible exponents, $\alpha\in [0.35,0.8]$, as shown in Fig.~\ref{fig:ratilde}(d).
This range includes the value 
$\alpha=1/2$ used by \cite{aubert2017}.
Yet, strikingly, the large values of $\alpha$ in this range are associated with a fast decrease of $\Em$ with $\Ek$; such $\alpha$ yield a minimum $\Em$ value that is too small for Earth's core (since the $\Ek_0$ and $\Em_0$ have been fixed).
This points to the importance of further pursuing this (very demanding) parameter survey.

%%%%%%%%%%%%%%%%
\section{Discussion}

We have tracked the weak-field and strong-field branches as the effects of rotation were increased (via decreasing Ekman number). The emergence (or not) of the strong-field branch is dependent on the value of the magnetic Ekman number. A sufficiently low $\Em$ (or high $\Pm$) is required for the strong-field branch to be realised. 
The critical value of the magnetic Ekman number, $\Em_c$, below which strong-field solutions are found has been found to scale with the Ekman number as $\Em\propto\Ek^\alpha$ 
where $\alpha \in[0.35, 0.8]$.
On the strong-field branch, one would expect an approximately $O(1)$ value for the Elsasser number given the balance of forces is MAC at large lengthscales and MC at a range of intermediate and small scales \citep{schwaiger2019,teed2023}.
Forming the ratio of the solenoidal forces (i.e.~curls of forces) accurately captures the effects of the Lorentz force across a range of rotation rates. We therefore advocate for a `Curlsasser number' as a {new} measure for identifying solutions as strong-field candidates.
Further work studying strong-field solutions in other parts of parameter space is obviously needed; for example, \cite{jones2025} recently suggested varying $q$ may be a fruitful endeavour for obtaining solutions with Earth-like dipole reversals.
In this work, we have cemented the importance of the strong-field branch (and bistability between dipolar branches) and demonstrated its persistence {and scaling} as input parameters are gradually moved towards realistic values for the Geodynamo.

%\acknowledgments
\section*{{\small Acknowledgments}}

This work was supported by the Science and Technology Facilities Council [grant number: ST/Y00146X/1] and DiRAC Project ACTP245.
The authors are grateful for time on the HPC resources of 
%the following two computing resources, which supported the numerical work for this study.
%\begin{itemize}
%    \item 
the DiRAC Memory Intensive service 
% TO BE REPLACED WITH THE LONG VERSION
%(Cosma8 / Cosma7 / Cosma6 [*]) 
%\textcolor{gray}{\sout{at Durham University, managed by the Institute for Computational Cosmology on behalf of the STFC DiRAC HPC Facility. %(www.dirac.ac.uk). 
%The DiRAC service at Durham}}, 
%funded by BEIS, UKRI and STFC capital funding, Durham University and STFC operations grants. DiRAC is part of the UKRI Digital Research Infrastructure.
%This work was also granted access to the HPC resources of MesoPSL financed
%by the Region Ile de France and the project Equip@Meso (reference
%ANR-10-EQPX-29-01) of the programme Investissements d’Avenir supervised
%by the Agence Nationale pour la Recherche.
and MesoPSL.
%Enter acknowledgments here. This section is to acknowledge funding, thank colleagues, enter any secondary affiliations, and so on.

%\bibliography{ enter your bibtex bibliography filename here }

\bibliographystyle{chicago} %%%% .BST file
\bibliography{references}

\begin{thebibliography}{}

\bibitem[\protect\citeauthoryear{Aubert, Gastine, and Fournier}{Aubert et~al.}{2017}]{aubert2017}
Aubert, J., T.~Gastine, and A.~Fournier (2017).
\newblock Spherical convective dynamos in the rapidly rotating asymptotic regime.
\newblock {\em Journal of Fluid Mechanics\/}~{\em 813}, 558--593.

\bibitem[\protect\citeauthoryear{Buffett}{Buffett}{2000}]{buffett2000}
Buffett, B. (2000).
\newblock Earth's core and the {G}eodynamo.
\newblock {\em Science\/}~{\em 288\/}(5473), 2007--2012.

\bibitem[\protect\citeauthoryear{Christensen and Aubert}{Christensen and Aubert}{2006}]{christensen2006}
Christensen, U. and J.~Aubert (2006).
\newblock Scaling properties of convection-driven dynamos in rotating spherical shells and application to planetary magnetic fields.
\newblock {\em Geophys. J. Int.\/}~{\em 166\/}(1), 97--114.

\bibitem[\protect\citeauthoryear{Christensen, Olson, and Glatzmaier}{Christensen et~al.}{1999}]{christensen1999}
Christensen, U., P.~Olson, and G.~Glatzmaier (1999).
\newblock Numerical modelling of the geodynamo: a systematic parameter study.
\newblock {\em Geophys. J. Int.\/}~{\em 138\/}(2), 393--409.

\bibitem[\protect\citeauthoryear{Dormy}{Dormy}{2016}]{dormy2016}
Dormy, E. (2016).
\newblock Strong-field spherical dynamos.
\newblock {\em J. Fluid Mech.\/}~{\em 789}, 500--513.

\bibitem[\protect\citeauthoryear{Dormy}{Dormy}{2025}]{Dormy2025}
Dormy, E. (2025).
\newblock Rapidly rotating magnetohydrodynamics and the {G}eodynamo.
\newblock {\em Ann. Rev. Fluid Mech.\/}~{\em 57}, 335--362.

\bibitem[\protect\citeauthoryear{Dormy, Oruba, and Petitdemange}{Dormy et~al.}{2018}]{dormy2018}
Dormy, E., L.~Oruba, and L.~Petitdemange (2018).
\newblock Three branches of dynamo action.
\newblock {\em Fluid Dynamics Res.\/}~{\em 50\/}(1), 011415.

\bibitem[\protect\citeauthoryear{Guseva, Petitdemange, and Tobias}{Guseva et~al.}{2025}]{Guseva2025}
Guseva, A., L.~Petitdemange, and S.~Tobias (2025).
\newblock Run-away transition to turbulent strong-field dynamo.
\newblock {\em J. Geophys. Res. Planets\/}~{\em 130}, e2024JE008496.

\bibitem[\protect\citeauthoryear{Horn and Aurnou}{Horn and Aurnou}{2022}]{horn2022}
Horn, S. and J.~M. Aurnou (2022).
\newblock The {E}lbert range of magnetostrophic convection. i. linear theory.
\newblock {\em Proc. R. Soc. A\/}~{\em 478\/}(2264), 20220313.

\bibitem[\protect\citeauthoryear{Horn and Aurnou}{Horn and Aurnou}{2025}]{horn2025}
Horn, S. and J.~M. Aurnou (2025).
\newblock The {E}lbert range of magnetostrophic convection. ii. comparing linear theory to nonlinear low-{R}m simulations.
\newblock {\em Proc. R. Soc. A\/}~{\em 481\/}(2310), 20240016.

\bibitem[\protect\citeauthoryear{Jones and Tsang}{Jones and Tsang}{2025}]{jones2025}
Jones, C.~A. and Y.-K. Tsang (2025).
\newblock Low inertia reversing geodynamos.
\newblock {\em Phys. Earth Planet. Inter.\/}~{\em 360}, 107303.

\bibitem[\protect\citeauthoryear{Kutzner and Christensen}{Kutzner and Christensen}{2002}]{Kutzner02}
Kutzner, C. and U.~Christensen (2002).
\newblock From stable dipolar towards reversing numerical dynamos.
\newblock {\em Phys. Earth Planet. Inter.\/}~{\em 131}, 29--45.

\bibitem[\protect\citeauthoryear{Mannix, Ponty, and Marcotte}{Mannix et~al.}{2022}]{Mannix2022}
Mannix, P.~M., Y.~Ponty, and F.~Marcotte (2022).
\newblock Systematic route to subcritical dynamo branches.
\newblock {\em Phys. Rev. Lett.\/}~{\em 129}, 024502.

\bibitem[\protect\citeauthoryear{Mason, Guervilly, and Sarson}{Mason et~al.}{2022}]{mason2022}
Mason, S.~J., C.~Guervilly, and G.~R. Sarson (2022).
\newblock Magnetoconvection in a rotating spherical shell in the presence of a uniform axial magnetic field.
\newblock {\em Geophys. Astrophys. Fluid Dyn.\/}~{\em 116}, 458--498.

\bibitem[\protect\citeauthoryear{Menu, Petitdemange, and Galtier}{Menu et~al.}{2020}]{menu2020}
Menu, M., L.~Petitdemange, and S.~Galtier (2020).
\newblock Magnetic effects on fields morphologies and reversals in geodynamo simulations.
\newblock {\em Phys. Earth Planet. Inter.\/}~{\em 307}, 106542.

\bibitem[\protect\citeauthoryear{Morin and Dormy}{Morin and Dormy}{2009}]{Morin09}
Morin, V. and E.~Dormy (2009).
\newblock The dynamo bifurcation in rotating spherical shells.
\newblock {\em Int. J. Mod. Phys. B\/}~{\em 23}, 5467--5482.

\bibitem[\protect\citeauthoryear{Schwaiger, Gastine, and Aubert}{Schwaiger et~al.}{2019}]{schwaiger2019}
Schwaiger, T., T.~Gastine, and J.~Aubert (2019).
\newblock Force balance in numerical geodynamo simulations: a systematic study.
\newblock {\em Geophys. J. Int.\/}~{\em 219}, S101--S114.

\bibitem[\protect\citeauthoryear{Soderlund, King, and Aurnou}{Soderlund et~al.}{2012}]{soderlund2012}
Soderlund, K., E.~King, and J.~Aurnou (2012).
\newblock The influence of magnetic fields in planetary dynamo models.
\newblock {\em Earth Planet. Sci. Lett.\/}~{\em 333}, 9--20.

\bibitem[\protect\citeauthoryear{Soderlund, Sheyko, King, and Aurnou}{Soderlund et~al.}{2015}]{soderlund2015}
Soderlund, K., A.~Sheyko, E.~King, and J.~Aurnou (2015).
\newblock The competition between lorentz and coriolis forces in planetary dynamos.
\newblock {\em Prog. Earth Planet. Sci.\/}~{\em 2\/}(1), 1--10.

\bibitem[\protect\citeauthoryear{Teed and Dormy}{Teed and Dormy}{2023}]{teed2023}
Teed, R. and E.~Dormy (2023).
\newblock Solenoidal force balances in numerical dynamos.
\newblock {\em J. Fluid Mech.\/}~{\em 964}, A26.

\bibitem[\protect\citeauthoryear{Willis, Sreenivasan, and Gubbins}{Willis et~al.}{2007}]{willis2007}
Willis, A., B.~Sreenivasan, and D.~Gubbins (2007).
\newblock Thermal core--mantle interaction: exploring regimes for ‘locked’ dynamo action.
\newblock {\em Phys. Earth Planet. Inter.\/}~{\em 165\/}(1-2), 83--92.

\end{thebibliography}

\end{document}